# TENSOR ROBUST PRINCIPAL COMPONENT ANALYSIS OF LIGHTNING IMAGES: BUTTERFLY EFFECT OF BLACK HOLES

[1]M. F. Yilmaz, [2]B. Karlik and [3]F. Yilmaz

[1]Independent Researcher, Fremont, CA, USA

[2] Neurosurgical Simulation Research and Training Centre, Department of Neurosurgery, Montreal Neurological, Institute and Hospital, McGill University, CANADA

[3] Department of Computer Sciences, New Jersey Institute of Technology, New Jersey, NY, USA

Tensor robust principal component analysis (robust PCA) has been applied to the lightning images. Robust PCA aims to classify the images into low-rank and sparse components. The low rank and sparse components correspond to static background separation and the dynamic (lightning) part of the images correspondingly. After classification, Singular Value Decomposition (SVD) unfold technique has been applied to sparse tensor, which transforms the tensor to spatial-temporal spaces in the form of vector matrixes. Spectral evolution shows the evolution of the polarization of the UV-Vis spectra. The contour maps of 2D energy density plots reveal the zero-point energy fluctuations of the bosons, fermions, and virtual matters from early to late stages. Detection of such fluctuations in the early stages can help to remote sensing of the lightning. 3D vector(ether) field representation of the sparse tensor of the brightest lightning reveals the butterfly effect of the black holes with the signatures of flapping. There is a correlation between the approximate entropy and the chaos trends of the 3D vector field analyzed by the Lyapunov exponent. In the brightest frame, the fractal dimension measures 0.8004, which is close to the Fibonacci Hamiltonian.

## I. INTRODUCTION

Lighting is regarded as one of nature's most fascinating and spectacular phenomena. However, it is relevant to note that there are still several intriguing facts about lightning that are not well understood. It has taken decades for scientists to investigate the most basic questions regarding lightning's origin and propagation. Technological advancements have provided new insights into the physics of lightning in recent years. A comprehensive review of the research on the lightning phenomenon can be found in the work of J. R. Dwyer and M. A. Uman et al 2014. As a result of the review of R. Dwyer and M. A. Uman, the phenomenon of lighting encompasses many branches of physics, such as atmospheric physics, plasma physics, and quantum electrodynamics, as well as posing a wide range of challenges that have yet to be resolved [1].

*Corresponding author: fthyilmaz53@gmail.com*

On the other hand, the data emerging from new technologies is very large and requires big data analysis. Dimensional reduction and future extraction are crucial parts of the analysis of big data. There is a possibility that conventional approaches, such as principal component analysis (PCA), linear discriminant analysis (LDA), independent component analysis (ICA), etc., might not be sufficient because covariant information may be lost [2 and 3]. In fact, the real data itself is naturally represented as multidimensional arrays or so-called tensors. In this regard, the recent advances in pattern recognition and, machine learning utilizing the tensors in big data analysis can also provide new aspects. For that reason, one must evaluate the multi-dimensional structure of the data with better perspectives of the features which can discover the hidden structures with a deeper understanding [4 and 5].

Although tensors and their decompositions were originally published in 1927, they were not well understood by the computer science community until the late 20th century. The advancement of computing power and a better understanding of multilinear algebra have allowed tensors to be applied to a variety of other fields, including statistics, data science, and machine learning [6]. It is important to note that many of the subjects of physics also naturally generate large amounts of multi-dimensional arrays or so-called tensors. For instance, the metric tensor plays a key role in apprehending the geometric and causal structure of spacetime in the theory of relativity. The electromagnetic field tensor describes the EM field with transformations in spacetime in the classical theory of electrodynamics. Tensor networks have been successfully utilized to represent product states of quantum many-body systems. For these reasons, the recent developments in tensor decompositions algorithms and tensor-based numerical methods have brought big attention in the physics community and overtaken the previously preferred methods [7 and 8].

Additionally, Yilmaz et al. have shown in numerous works that linear and probabilistic PCA and LDA can be used to efficiently deduce many features of physics from spectral data of plasmas at low and high temperatures. As an example, spectra of high energy density K-shell Al and L-shell Mo plasmas have revealed self-organized quantized V-shaped pairing structures in the presence of the electron beams. The LDA analysis of K-shell Al plasma spectral data indicated the existence of Langmuir turbulence structures and that such turbulence correlated with the predator-prey analogy between electron and ion oscillations. B. Karlik and Yilmaz et al.,2020 reported that the 3D vector (ether) fields obtained by PCA and LDA over UV spectra of gold nanoparticles revealed strange attractors and quantum confinement of collective plasmonic oscillations [9,10,11,12 and 13].

It has been demonstrated by Yilmaz et al.2021 that Kernel PCA over spectra of Argonne discharge plasmas can reveal spatial structures such as strange attractors. Further, Yilmaz has used the Welch transfer methods in order to describe the transformation from a low-energy-density vector field to a higher-density vector field. Based on the results, it appears that such a transformation was achieved in the form of zero-point energy fluctuations between the particles, antiparticles, virtual particles, and virtual antiparticles [14]. In contrast, the tensor model naturally describes the transformation of vector fields without the need for any additional transfer functions. In this work, the energy density evolution of an incident lightning flash has been studied using the tensor decomposition of images taken just before and after the incident. A study of the degree of chaos in 3D vector fields has been conducted using the Lyapunov exponent, approximate entropy, and the fractal dimension. The evolution of the vector fields has also been viewed from a topological perspective.

## II. ROBUST PCA OF LIGHTNING IMAGES

Natural-source image data are multidimensional, even if they are processed as two-dimensional matrices. Multidimensional data is typically represented as matrices due to their usefulness and solid theoretical basis. In this context, a variety of tensor-based approximations have been explored, including general tensor discriminant analysis, tensor adaptations of classical principal component analyses, and multilinear principal component analyses. Using these techniques, image data is often preserved in its natural form, no information is lost, and better recognition rates are achieved. In addition to principal component analysis (PCA), robust principal component analysis (robust PCA) can also be classified as a low-rank or a low-rank/sparse decomposition approach. Using tensor robust principal component analysis, one can solve low-rank and sparse decomposition, which is particularly useful when dealing with shallowly buried objects and rough surfaces [15,16,17 and 18].

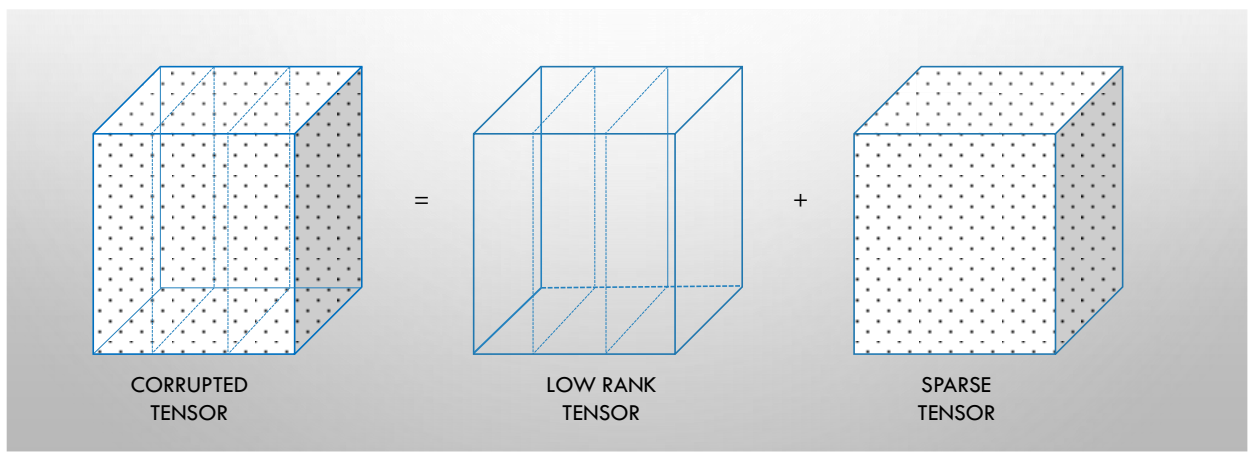

Figure.1: Illustrations of Robust PCA (up row) and our Tensor Robust PCA (bottom row). Robust PCA: low-rank and sparse matrix decomposition from noisy matrix observations. Tensor Robust PCA: low-rank and sparse tensor decomposition from noisy tensor observations.

Object detection and outlier detection are some of the applications that Tensor Robust PCA is widely used in, as well as background subtraction and denoising. Decomposition of tensors using robust PCA involves partitioning them into low-rank and sparse components. A sparse tensor represents moving objects in video images, whereas a low-rank tensor represents stationary objects. A tensor of low rank that has been severely corrupted is recovered by the sparse component. This feature of Robust PCA assists video surveillance systems in detecting foreground objects by initializing the background model [15,16,17 and 18]. The evolution of the lightning and darkening with low rank and sparse versions are illustrated in Fig.1 and Fig.2 correspondingly.

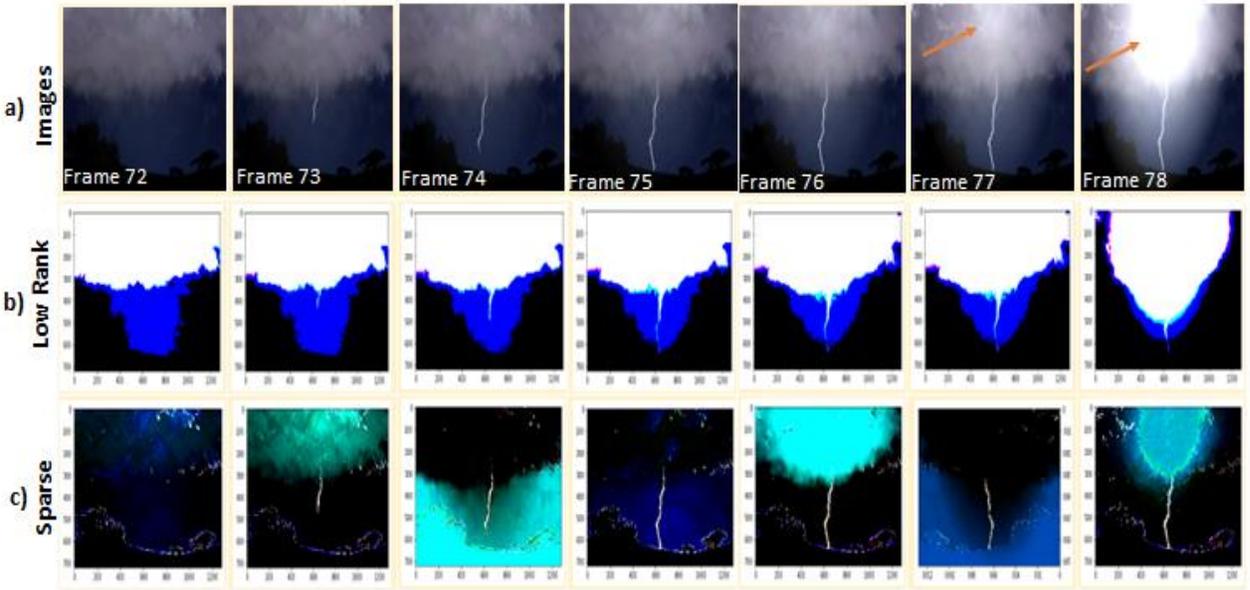

Figure 1. a) Images of lightning b) Low-rank c) Sparse version of the images

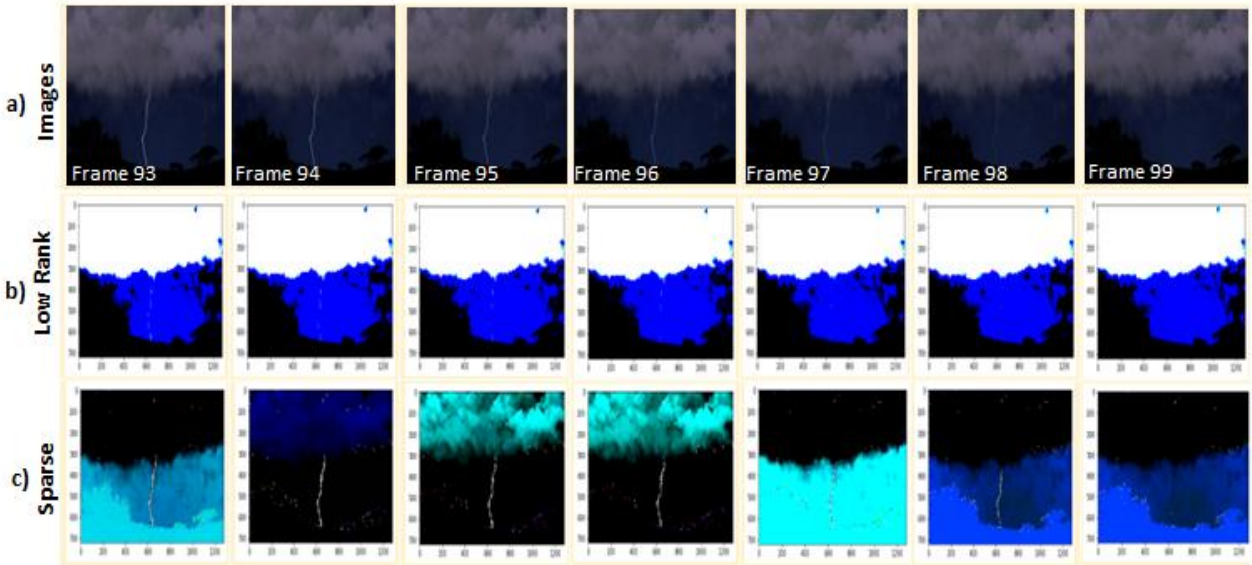

Figure 2. a) Images of darkening b) Low-rank c) Sparse version of the images

## III. DISCUSSION

In Fig.3 1D,2D, and 3D of vector spectra of Frame78 has been illustrated. Fig.3-a shows the UV spectrum, and its evolution of circular polarization (Stokes V) has been illustrated. Fig.3 b shows the diverging view of the 2D representation of the |PC1> and |PC2> spectra. In Fig3.c 3D scattering view of the vector spectra is illustrated. A 3D vector field (ether field) view reveals the butterfly effect of the black hole, along with a signature of flapping.

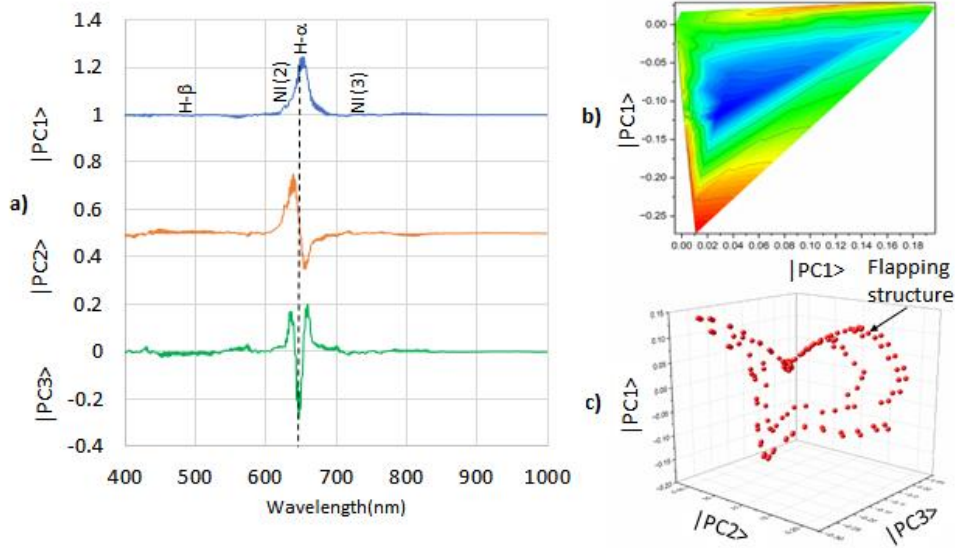

Figure 3. a) Vector spectra of |PC1>, |PC2> and |PC3>, b) topologic view of 2D representation of |PC1> versus |PC2>, c) 3D scattering view (|PC1> versus |PC2> versus |PC3>) of the vector spectra of the Frame78

Butterfly effects are often regarded as the epitome of chaotic behavior in dynamical systems: a system's dependence on its initial conditions. When a chaotic system is involved, even a slight change in initial conditions, such as the flapping of a butterfly's wings, can result in dramatic changes in the behavior of the system [19 and 20]. There is a great deal of knowledge about butterfly effects in relation to weather since they have been demonstrated in standard weather prediction models. In semiclassical and quantum physics, many phenomena depend sensitively on initial conditions, such as atoms in strong fields and anisotropic Kepler's problem [21].

Black holes, on the other hand, are thought to be responsible for many of the most energetic processes in the universe, making them one of the most fascinating objects in physics. There has been a great deal of research on understanding such energy processes. According to Stanford et al., 2014, black holes also exhibit the butterfly effect. Stanford has proposed that shock waves near a black hole's horizon are responsible for the chaotic behavior within thermal CFTs. A few quanta are thrown into the horizon, which corresponds to perturbed thermofield double states on the boundaries by local operators, causing local shock waves in the bulk to spread. By utilizing holography, butterfly effects can be explained by the spreading of shock waves at the horizon of boundary field theory. After the quick decay of scrambling time $t = \frac{\beta}{2\pi} \ln S$, where "S" represents the entropy of a black hole, the butterfly effect can be elucidated [22].

According to Moreta et al.,2021 butterfly effects with an emergent thermal nature could appear not only in black holes but also in various situations: the Unruh effect in accelerated observers, acoustic Hawking radiation in supersonic fluids, moving mirrors and so on. Moreta considers a classical dynamical system with a deterministic Lyapunov exponent of 0.01 which is non-thermal (*T=0*). As a result of quantizing this system, quantum fluctuations would be triggered to mimic thermal fluctuations with the temperature of $T \sim \frac{\hbar \lambda L}{2\pi} + O(\hbar^2)$ in order to establish the analog Hawking radiation. $O(\hbar^2)$ corresponds to quantum correction of the system due to inverse harmonic potential. So, the emergent thermal nature could be explained by inverse harmonic oscillations, which exhibit butterfly effects at the end of their oscillations [23].

In Fig4, a sparse version of the lightning images and their corresponding 2D topologic energy density and 3D vector field plots has been illustrated. The lighting gets brighter, and the main flash happens in frame 78. The 2D energy density contour map (|PC1> versus |PC2>) of frames 72. and 73. c) illustrates the convergence flow between the antimatter and virtual matter seas. In the antimatter sea, the fluctuations eventually converge to form well near the zero point, which can be viewed as a sign of a white hole [24]. It is well known that in quantum mechanics, zero-point energy (ZPE) refers to the lowest energy attainable by a system. According to Heisenberg's uncertainty principle, quantum systems are constantly fluctuating in their lowest energy states, unlike classical mechanics [25 and 26]. By using the Welch transfers method over the vector fields (ether fields) of the glow discharge Argonne plasma, Yilmaz et al., 2021 have been able to distinguish the zero-point energy fluctuations of bosons, fermions, and virtual matters [14]. According to the 2D maps, as the flow forms a convergence in the antimatter sea, an occurrence occurs in the matter sea, which can be explained by quantum entanglement. [27,28 and 29]. When the flash is present (Frames 77 and 78), the emergence of the particles in the matter sea is observed out of the black hole near the point greater than zero. In Figure 5, the darkening process has been illustrated. 2D topologic views show the convergence of the flow from the sea of matter and virtual matter towards to white hall in the antimatter sea. Finally, the flow diverges in the antimatter sea from the white hole. Also comparing the 3D vector field of Frame 72 and 99 shows the similarities, which can be the exhibition of the protection of the information. So, the 3D patterns we have obtained during the lightning and darkening can be very beneficial for the quantum sensing of lightning and lightning strikes [30 and 31].

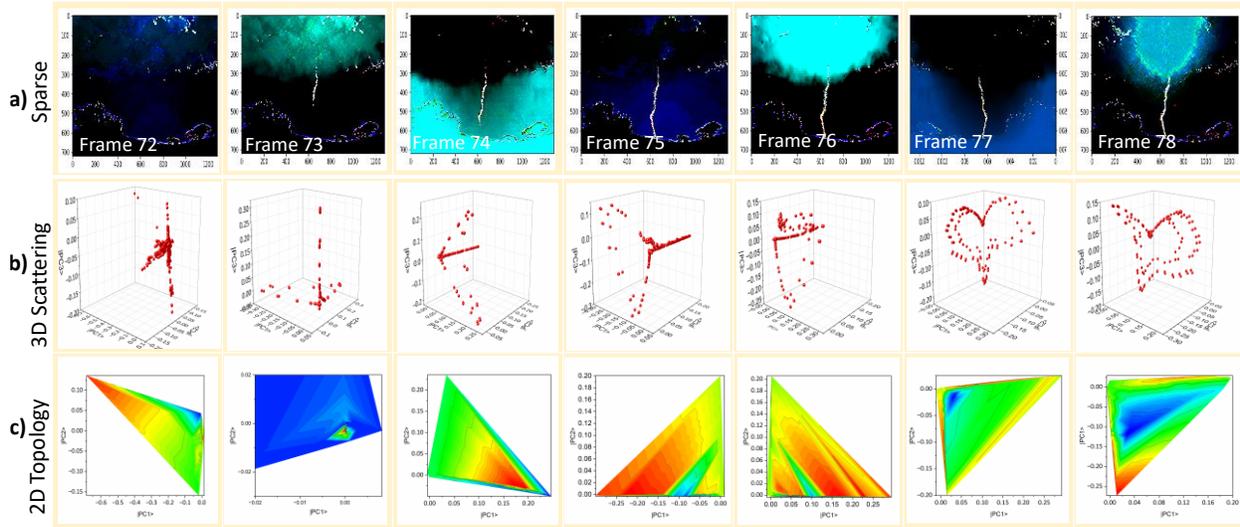

Figure 4. a) Sparse images of lightning process b) 3D vector fields c) topologic plots of 3D vector fields

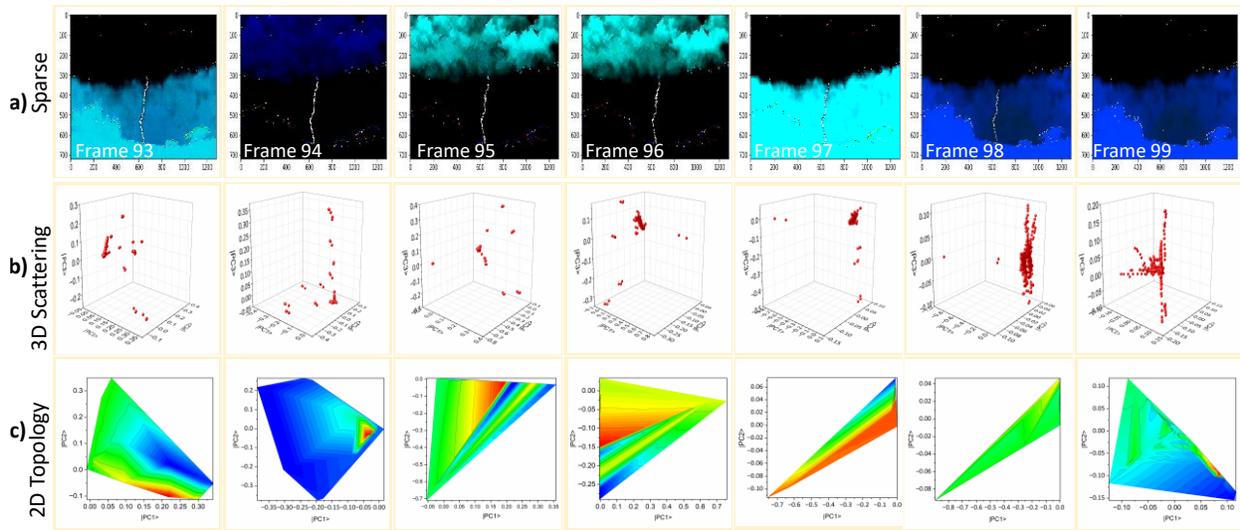

Figure 5. a) Sparse images of the darkening process b) 3D vector fields c) topologic plots of 3D vector fields

We have evaluated the chaos analysis of the 3D vector fields of the considered frames by using the Lyapunov exponent, approximate entropy, and fractal dimensions. It is well known that the Lyapunov characteristic exponent provides a measure of how quickly trajectory separation occurs between infinitesimally close points in a dynamical system**.** In the case of a Lyapunov exponent of zero, the system reaches some state of equilibrium, and orbits become fixed or are on the verge of becoming fixed as they approach a fixed point. It is generally considered that negative Lyapunov exponents are indicative of a dissipative system whose orbit is attracted to a stable fixed point or stable periodic orbit. Whenever the Lyapunov exponent exceeds zero, the orbit is characterized by chaos and instability and each neighborhood will be visited as part of the phase space [31].

Figures 6. a and d) show the trends of the chaos of lighting and darkening by the Lyapunov exponent. Trends show that the system becomes more chaotic towards lightening and darkening. Consequently, fluctuations emerge to matter sea (lightning) or antimatter sea (darkening), and they become more chaotic. A measure of approximation of entropy is based on the frequency of repeating patterns for evaluating the level of randomness involved in a data series. Its low values indicate that there are many repeated patterns, whereas its high values indicate that there are many unpredictable events [32]. Entropy measurements in Figures 6. b and e) also follow almost the same trends of the Lyapunov exponent except the frame 72. So, considering the systems consist of the quantum fluctuations of the boson, fermions, and virtual matter follows the second rule of thermodynamics [33 and 34]. Objects exhibiting the property of self-similarity are called fractals; they are geometric shapes that can be broken down into smaller parts with each smaller part being a miniature copy of the whole. As a statistical quantity, fractal dimension is described as the way in which a fractal occupies space [35]. Fractal dimensions of the vector fields are illustrated in Figures 6. c and f) and the figures show that the fractal dimension trends are slower when compared to the Lyapunov exponent and approximate entropy. Furthermore, Frame 78 with the butterfly exhibition has a fractal dimension of 0.80004. The spectrum of the Fibonacci Hamiltonian has a fractal dimension of 0.88. In the regime of large coupling, the Fibonacci Hamiltonian demonstrates upper and lower bounds for its fractal dimension [36].

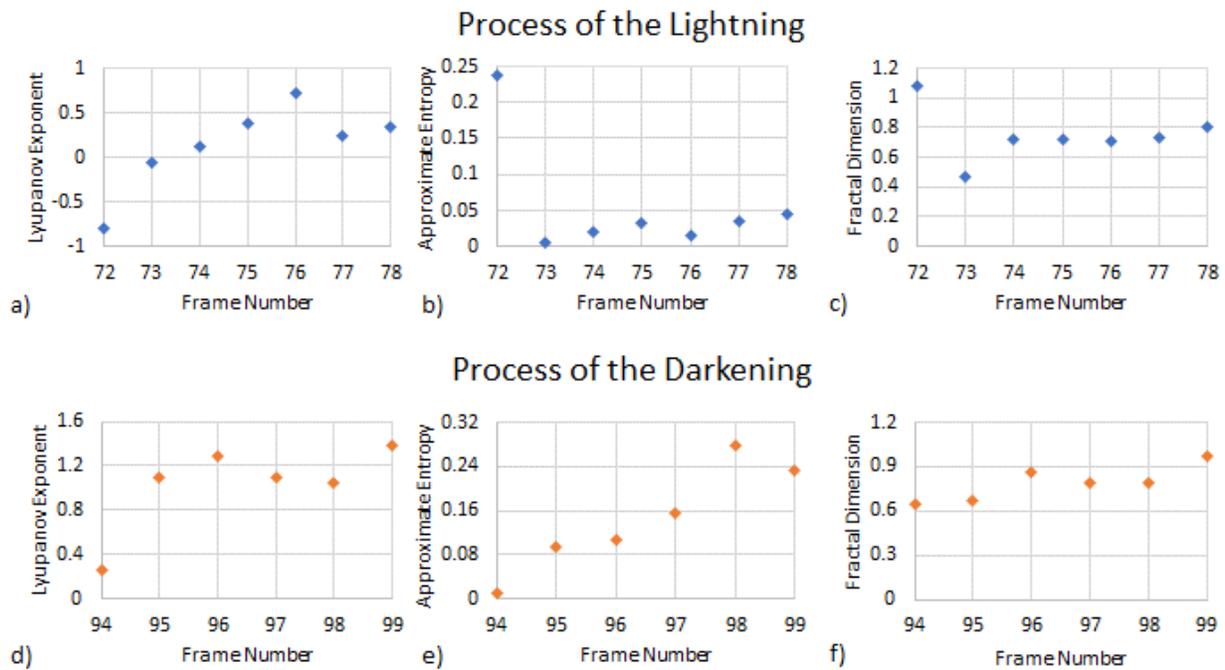

Figure 6. a and d) Lyapunov Exponent b and e) approximate entropy c and f) fractal dimension during the process of the lightening and darkening correspondingly.

## IV. CONCLUSIONS

Our results show that tensor robust PCA can be a powerful tool to study the quantum level study of lightning phenomena over image data. 1D representation of the robust PCA successfully elucidates the spectra of lightning with polarization structures. 2D energy density contour maps illustrate zero-point energy fluctuations without utilizing any transfer functions. According to Yilmaz et al., 2021, these fluctuations obtained by robust PCA of image data are similar to those obtained by Welch transfer of the vector fields of glow discharge Argonne plasma spectral data. Furthermore, 3D vector fields (ether fields) exhibit a butterfly effect in the black hole with the signatures of the flapping. Studying the early lightning and darkening processes using 3D vectors can serve as a test bed for deep learning algorithms designed to detect lightning in its earliest stages. The resemblance of the 3D vector field of very early and late stages can be the signature of the protection of quantum information. The chaos trends predicted by the Lyapunov exponent are in agreement with the approximate entropy of the data. The fractal dimension of the brightest frame has 0.8004 which is close to the Fibonacci Hamiltonian.

ACKNOWLEDGEMENTS

This work is dedicated to the victims of the wildfires that occurred in California in 2020 and Oregon in 2021. We would like to express our gratitude to Dr. John Luginsland of the US Air Force Office of Scientific Research for his support and enthusiasm throughout our research. This set of lightning images was obtained from www.istockphoto.com, with order number 2064365124.

LIST OF REFERENCES

1- Dwyer, J. R., & Uman, M. A. (2014). The physics of lightning. *Physics Reports*, *534*(4), 147-241.
2- Chebbi, I., Boulila, W., & Farah, I. R. (2015). Big data: Concepts, challenges and applications. *Computational collective intelligence*, 638-647.
3- Reddy, G. T., Reddy, M. P. K., Lakshmanna, K., Kaluri, R., Rajput, D. S., Srivastava, G., & Baker, T. (2020). Analysis of dimensionality reduction techniques on big data. *IEEE Access*, *8*, 54776-54788.
4- Baskaran, M., Meister, B., Vasilache, N., & Lethin, R. (2012, September). Efficient and scalable computations with sparse tensors. In *2012 IEEE Conference on High Performance Extreme Computing* (pp. 1-6). IEEE.
5- Huy, P. A. (2011). Algorithms for tensor decompositions and applications. *PhD Thesis, Kyushu Institute of Technology*.
6- Rabanser, S., Shchur, O., & Günnemann, S. (2017). Introduction to tensor decompositions and their applications in machine learning. *arXiv preprint arXiv:1711.10781*.
7- Khan, Q. (2017). Tensor Analysis and its Applications in Physics.
8- Ashtekar, Abhay, Gary T. Horowitz, and Anne Magnon-Ashtekar. "A generalization of tensor calculus and its applications to physics." *General Relativity and Gravitation* 14.5 (1982): 411-428.
9- Sadiq, N., & Ahmad, M. (2019). Kinetic Alfven waves in dense quantum plasmas with the effect of spin magnetization. Plasma Research Express, 1(2), 025007.

10- Yilmaz, M. F., Danisman, Y., Larour, J., & Aranchuk, L. (2015). Principal component analysis of electron beams generated in K-shell aluminum X-pinch plasma produced by a compact LCgenerator. High Energy Density Physics, 15, 43-48.
11- Yilmaz, M. F., Danisman, Y., Larour, J., & Arantchouk, L. (2019). Linear discriminant analysis based predator-prey analysis of hot electron effects on the X-pinch plasma produced Kshell Aluminum spectra. Scientific reports, 9(1), 1-8.
12- Karlik, B., Yilmaz, M. F., Ozdemir, M., Yavuz, C. T., & Danisman, Y. (2021). A Hybrid Machine Learning Model to Study UV-Vis Spectra of Gold Nanospheres. Plasmonics, 16(1), 147- 155.
13- Moldabekov, Z. A., Bonitz, M., & Ramazanov, T. S. (2019). A critique of" quantum dusty plasmas". arXiv preprint arXiv:1901.10839.
14- Yilmaz, M. F., Tinaztepe, R., Karlik, B., & Yilmaz, F. (2021). Welch's method transform of strange attractors of argon glow discharge plasmas: symmetric zero point energy fluctuations of particles, anti-particles and virtual particles. *arXiv preprint arXiv:2106.01493*.
15- Lu, C., Feng, J., Chen, Y., Liu, W., Lin, Z., and Yan, S. (2019). "Tensor Robust Principal Component Analysis with a New Tensor Nuclear Norm". IEEE Transactions on Pattern Analysis and Machine Intelligence, 42(4), 925-938.
16- Kumlu, D. and Gundogdu, B. (2021). "A Novel Tensor RPCA Method For Clutter Suppresion in GPR Images". Journal of Naval Sciences and Engineering,(1), 27-42.
17- Cordolino Sobral, A. (2018). "Robust low-rank and sparse decomposition for moving object detection: from matrices to tensors ". PHD Thesis, Universite de La Rochelle, https://tel.archives-ouvertes.fr/tel-01753784
18- Lu, C. (2021). Transforms based Tensor Robust PCA: Corrupted Low-Rank Tensors Recovery via Convex Optimization. In Proceedings of the IEEE/CVF International Conference on Computer Vision (pp. 1145-1152).
19- Lorenz, E. N. (1963). Deterministic nonperiodic flow. *Journal of atmospheric sciences*, *20*(2), 130-141.
20- Hilborn, R. C. (2000). *Chaos and nonlinear dynamics: an introduction for scientists and engineers*. Oxford University Press on Demand.
21- Keppler, M., Benisty, M., Müller, A., Henning, T., Van Boekel, R., Cantalloube, F., ... & Weber, L. (2018). Discovery of a planetary-mass companion within the gap of the transition disk around PDS 70. *Astronomy & Astrophysics*, *617*, A44.
22- Stanford, D. (2014). *Black holes and the butterfly effect*. Stanford University.
23- Morita, T. (2021). Analogous Hawking radiation in butterfly effect. *SciPost Physics Proceedings*, (4), 007.
24- Bianchi, E., Christodoulou, M., d'Ambrosio, F., Haggard, H. M., & Rovelli, C. (2018). White holes as remnants: a surprising scenario for the end of a black hole. *Classical and Quantum Gravity*, *35*(22), 225003.
25- Milton, K. A. (2001). *The Casimir effect: physical manifestations of zero-point energy*. World Scientific.
26- Maggiore, M. (2011). Zero-point quantum fluctuations and dark energy. *Physical Review D*, *83*(6), 063514.
27- Aspect, A. (1999). Bell's inequality test: more ideal than ever. *Nature*, *398*(6724), 189-190.
28- Clauser, J. F., & Shimony, A. (1978). Bell's theorem. Experimental tests and implications. *Reports on Progress in Physics*, *41*(12), 1881.
29- Fickler, R., Lapkiewicz, R., Plick, W. N., Krenn, M., Schaeff, C., Ramelow, S., & Zeilinger, A. (2012). Quantum entanglement of high angular momenta. *Science*, *338*(6107), 640-643.

30- Cummins, K. L., Murphy, M. J., & Tuel, J. V. (2000, September). Lightning detection methods and meteorological applications. In *IV International Symposium on Military Meteorology* (pp. 26-28).
31- Christian, H. J. (1999, June). Optical detection of lightning from space. In *11th International Conference on Atmospheric Electricity*.
32- Wolff, Rodney CL. "Local Lyapunov exponents: looking closely at chaos." *Journal of the Royal Statistical Society: Series B (Methodological)* 54.2 (1992): 353-371.
33- Lent, C. S. (2019). Information and entropy in physical systems. In *Energy Limits in Computation* (pp. 1-63). Springer, Cham.
34- Delgado-Bonal, A., & Marshak, A. (2019). Approximate entropy and sample entropy: A comprehensive tutorial. *Entropy*, *21*(6), 541.
35- Freistetter, F. (2000). Fractal dimensions as chaos indicators. *Celestial Mechanics and Dynamical Astronomy*, *78*(1), 211-225.
36- Damanik, D., Embree, M., Gorodetski, A., & Tcheremchantsev, S. (2008). The fractal dimension of the spectrum of the Fibonacci Hamiltonian. *Communications in mathematical physics*, *280*(2), 499-516.